\documentclass{article}  
\usepackage{breckenr2005}
\frompage{000} \topage{000}                                              
\def\rightmark{Heavy flavor production at $\sqrt{s_{NN}} = 200\ GeV$}
\def\leftmark{Y. Kwon et al.}
 
\title{
Open heavy flavor production from light ion collisions \\
\center  
at RHIC
} 
\authors{
{Y. Kwon$^1$ for the PHENIX collaboration}\\[2.812mm]
{\normalsize
\hspace*{-8pt}$^1$Dept. of Physics, Univ. of Tennessee, 
Knoxville, TN 37996, USA\\[0.2ex] 
}
}
 
\abstract{
PHENIX measures leptons at mid and forward rapidities
and 
extracts leptons resulting from semi-leptonic decays 
of heavy quarks.
We present PHENIX results related to heavy quark production, 
specifically 
the invariant cross section of the non-photonic single electrons
produced at midrapidity 
in {\it p+p} and {\it d+Au} collisions at $\sqrt{s_{NN}} = 200\ GeV$.
The measured cross section for {\it p+p} collisions is compared
with the NLO prediction, and a possible excess over the prediction is noted.  
The measured cross section for {\it d+Au} collisions scales
with the number of colliding nucleon pairs 
from the {\it p+p} spectra.
}

 
\begin{document}
 
\maketitle
\setcounter{page}{1}

\section{Introduction}\label{sec:introduction}

The parton model and 
{\bf p}erturbative {\bf Q}uantum {\bf C}hromo{\bf D}ynamics ( {\bf pQCD} )
can be applied to the violent strong interaction processes 
characterized by a large scale~\cite{pub:factor}. 
As an example, it is applied to the hard particle production 
where $p_{T}$ serves as the large scale.
Neutral pion production in {\it p+p} collisions at $\sqrt{s} = 200\ GeV$
in the range $2 < p_{T} < 13\ GeV/c$ measured by the PHENIX 
experiment at RHIC can be well described 
by the {\bf N}ext-to-{\bf L}eading-{\bf O}rder ({\bf NLO}) 
pQCD~\cite{pub:pi0phenix}.

Heavy quark production belongs to the category 
of violent strong interaction processes
due to its large mass.
While classification of the charm quark as a heavy quark 
is not yet fully settled~\cite{pub:factor}, 
the parton model and pQCD 
are often applied to the charm quark production.
Within the given limitations,
predictions for charm hadro-production exist 
from lower collision energies 
including the Fermi Lab fixed target program~\cite{pub:charm2}.

Further predictions were made for heavy quark production 
in the ion collisions.  
Systematic studies of charm production in {\it p+p} and p+nucleus collisions 
have been proposed as a sensitive way to measure the parton distribution
function in nucleons, and nuclear shadowing effects~\cite{pub:pdf}.
For heavy ion collisions at RHIC energies, 
heavy quark energy loss~\cite{pub:heavyloss},
modification of charm quark hadronization~\cite{pub:charmcoal},
in-medium effects on open- and hidden-charm states~\cite{pub:inmedium},
possible $J/\psi$ suppression~\cite{pub:jpsisup},
and charm flow~\cite{pub:charmflow}
have been proposed as possibilities.

\section{Theoretical background}\label{sec:motivation}  
\label{sec:theory}

Factorization of the hadro-production cross section
of heavy flavored hadrons 
is described as follows.
The differential cross-section for the production of 
a heavy flavored particle
(a charmed particle in the particular case), H, 
in the collision of two hadrons, A and B, is believed to factorize 
in the following manner~\cite{pub:factor},

\begin{equation}
d\sigma [A+B \rightarrow H+X] = 
\sum_{ij} f_{i/A} \otimes f_{j/B} \otimes
d\hat{\sigma} [ij \rightarrow h\bar{h}+X] 
\otimes D_{h \rightarrow H} + remainder 
\label{eq:factorization}
\end{equation}
 
Here i,j denote partons which are point-like, 
$f_{i/A}$ and $f_{j/B}$ are parton distribution functions, and
$D_{h \rightarrow H}$ is the fragmentation function 
for the heavy quark h hadronizing into H.
The parton distribution functions and the fragmentation functions 
are process independent.
The $d\hat{\sigma} [ij \rightarrow h\bar{h}+X]$
is a perturbatively calculable short-distance cross section.
The remainder represents corrections
to the factorized form of the cross section that are 
power-suppressed by $\Lambda_{QCD}/m_{h}$ 
or 
possibly $\Lambda_{QCD}/p_{T}$ if $p_{T} \gg m_{h}$. 
A large quark mass $m_{h}$ makes the remainder smaller,
and
application of pQCD to the factorized term becomes possible.
 
As the collision energy increases, 
the factorized term increases faster than the remainder, and
the hard processes (factorized term) gain importance 
at high collision energies.
We can test the validity of the factorization at RHIC collision energies
by measuring heavy flavored hadron production 
for {\it p+p} collisions. 

The factorization scheme has further consequences 
for the collisions of ions.
For the interaction between point-like partons,
we can approximate the nuclear parton distribution as follows.
\begin{eqnarray}
f_{i/Au} \approx 79\ f_{i/p} + 118\ f_{i/n} \approx 197\ f_{i/N}, \nonumber 
\end{eqnarray}
and
\begin{eqnarray}
f_{i/d} \approx f_{i/p} + f_{i/n} \approx 2\ f_{i/N} \nonumber
\end{eqnarray}
where $f_{i/N}$ is the parton distribution function inside the nucleon. 

If the charm producing processes are point-like and
there is no modification of the initial parton distribution
for the {\it d + Au} collisions,
the charm production cross section will scale with $2 \times 197$,
i.e. the number of colliding nucleon pairs. 
Surprisingly, 
this extension does not work for high $p_{T}$ particles 
produced in central Au+Au collisions~\cite{pub:PHENIXquench},
and 
careful studies of the stated scaling are desirable
for the various hard processes.

\section{Measurements}\label{sec:measurement}

\begin{figure}[htb]
\vspace*{-1.5cm}
\epsfxsize=3.0in
\insertplot{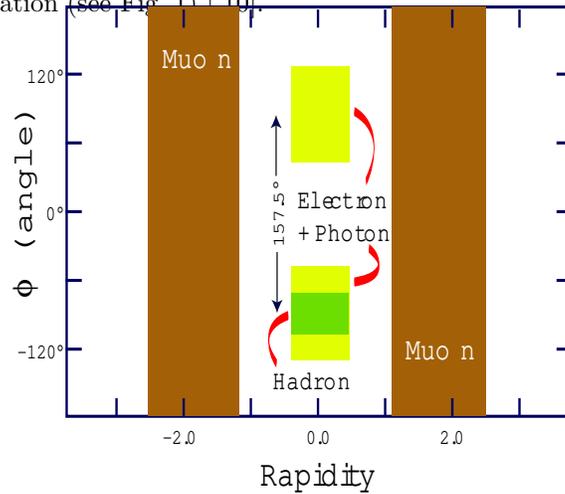}
\vspace*{-0.8cm}
\caption[]{Acceptance of the PHENIX experiment.}
\label{fig:acceptance}
\end{figure}

PHENIX is capable of measuring leptons at mid and forward rapidities.
PHENIX uses global detectors to characterize the collisions,
a pair of central spectrometers at mid rapidity to measure electrons,
hadrons, and photons, and a pair of forward spectrometers to measure
muons.
Each spectrometer has a large geometric acceptance of about one
steradian, excellent energy and momentum resolution, and 
particle identification 
(see Fig.~\ref{fig:acceptance})~\cite{pub:detphenix}. 

The PHENIX central arms consists of tracking systems for charged particles
and electromagnetic calorimetry.
The calorimeter is the outermost subsystem of the central arms
and provides measurements of both photons
and energetic electrons.
The tracking system uses three sets of Pad Chambers 
(PC) to provide 
precise three-dimensional space points needed for pattern recognition.
The precise projective tracking of the Drift Chamber 
(DC) is the basis
of the excellent momentum resolution.
A Time Expansion Chamber (TEC) in the east arm
provides additional tracking and particle identification.
The Time-of-Flight (ToF) 
and 
Ring-Imaging CHerenkov (RICH) detectors
provide particle identification.
The RICH provides separation of electrons 
from the large number of copiously produced pions. 
Using information from the RICH, the TEC, 
and the electromagnetic calorimeter 
it is possible to reject pion contamination of identified electrons 
to one part in $10^{4}$ over a wide range of momenta.
 
Each forward spectrometer is based on a muon tracker 
inside a radial magnetic field followed by a muon identifier, 
both with full azimuthal acceptance.
The muon trackers consist of 
three stations of multiplane drift chambers that provide precision tracking.
The muon identifier consist of 
alternating layers of steel absorbers 
and 
low resolution tracking layers of streamer tubes of the Iarocci type.
With this combination, 
the pion contamination of identified muons is typically $3 \times 10^{-3}$,
and 
can be estimated and subtracted statistically.
Measurement of {\it prompt muons} resulting from semi-leptonic decays
of heavy flavored hadrons is also possible,
and analysis is actively persued.

\begin{figure}[htb]
\vspace*{4.5cm}
\epsfxsize=3.5in
\insertplot{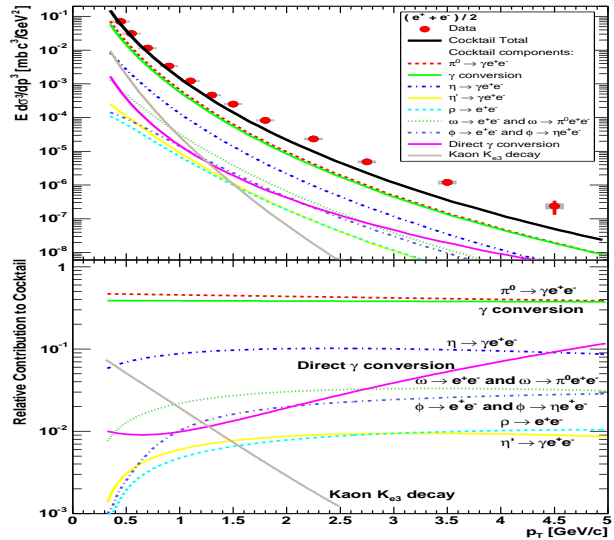}
\vspace*{-0.8cm}
\caption[]{
The inclusive single electron spectra 
and 
composition of backgrounds, 
{\it p+p} collisions at $\sqrt{s_{NN}} = 200\ GeV$.
}
\label{fig:cocktail}
\end{figure}

PHENIX 
measures the non-photonic single electron spectra by first measuring 
the inclusive electron spectra 
and then subtracting the contribution from photonic sources.
Inclusive electrons contain two components:
(1) {\it non-photonic} - primarily semi-leptonic decays of mesons
containing heavy (charm and bottom) quarks, and 
(2) {\it photonic} - Dalitz decays of light neutral mesons 
($\pi^{0}, \eta, \eta^{\prime}, \rho, \omega,$ and $\phi$)
and photon conversions in the detector material.
The PHENIX experiment uses two approaches, 
{\it converter subtraction} and {\it cocktail subtraction},
to estimate the photonic component.  

The converter subtraction is a data-driven approach.
A photon converter 
(a thin brass tube of 1.7\% radiation length thick) 
is installed in the middle of data-taking.
The photon converter multiplies 
the photonic contribution to the electron yield in a well-defined manner
and 
hence the photonic component can be estimated.

The cocktail subtraction simulates electron production 
from the known sources and subtract them from the inclusive spectra.
PHENIX has measured the $p_T$ distributions 
of $\pi^{\pm}$ and $\pi^{0}$.
We fit a power law function to the combined data sets 
to determine the input $\pi^{0}$ spectrum for the decay generator.
The $p_T$ distribution of any other hadrons is 
obtained from the $\pi^{0}$ spectrum 
by replacing $p_T$ with $\sqrt{p_T^2 + m_{h}^{2} - m_{\pi^0}^2}$.
In this parameterization
$h/\pi^0$ ratios approach constants at high $p_T$.
The asymptotic ratios used to fix the relative normalizations
are determined from the world data including the PHENIX results.
Photon conversions are evaluated using a combination
of the GEANT simulation and the hadron decay generator.

The converter subtraction and the cocktail subtraction
yield consistent results, 
and
Fig.~\ref{fig:cocktail} shows the inclusive electron spectra 
and composition of the estimated background.

\section{Result}\label{sec:Result}  


\begin{figure}[h]
\vspace*{4.5cm}
\epsfxsize=4.0in
\insertplot{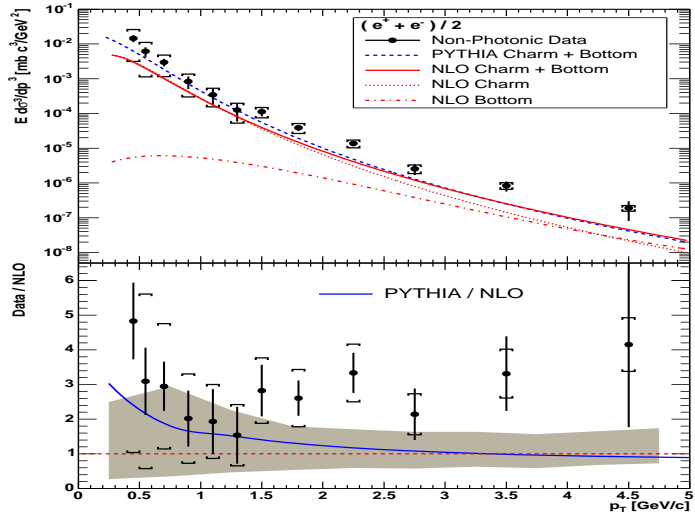}
\vspace*{-0.8cm}
\caption[]{
The non-photonic single electron spectra from the {\it p+p} collisions 
at $\sqrt{s_{NN}} = 200\ GeV$
and comparison with the NLO pQCD prediction.}
\label{fig:nlo}
\end{figure}


The cocktail subtracted invariant electron cross section 
is shown on Fig.~\ref{fig:nlo}.
The PYTHIA calculation  
($m_{c} = 1.25\ GeV/c^{2}$, $m_{b} = 4.1\ GeV/c^{2}$,
$K = 3.5$, and $< k_{t} > = 1.5)$)
is displayed on the same figure.
The NLO pQCD prediction of heavy flavor-related electrons
was made using the HVQLIB package.
The heavy quark production cross section was calculated 
using HVQLIB and PYTHIA, 
and the ratio of heavy quark yields 
$(dN_{q}/dp_{T})_{HVQLIB}$ to $(dN_{q}/dp_{T})_{PYTHIA}$
was used as a weight for the PYTHIA decay electrons.
This method enables us to use an exact NLO pair production
calculation together with fragmentation and decay kinematics
from PYTHIA.
Within the NLO pQCD calculation,
charm decays are the major source of non-photonic single electrons
for most of the observed $p_{T}$ range. 
The NLO prediction for open charm yields harder spectra 
than the PYTHIA calculation, 
but 
an excess of data over both predictions is seen beyond intermediate $p_{T}$.
An elaborate theoretical study has appeared recently 
and 
also shows the observed excess~\cite{pub:FONLL}.
The uncertainty of theoretical prediction 
mostly comes from the mass and the scale uncertainties,
and is relatively large. 
Reduced experimental errors will contribute to a definitive statement.

\begin{figure}[htb]
\vspace*{.8cm}
\epsfxsize=4.0in
\insertplot{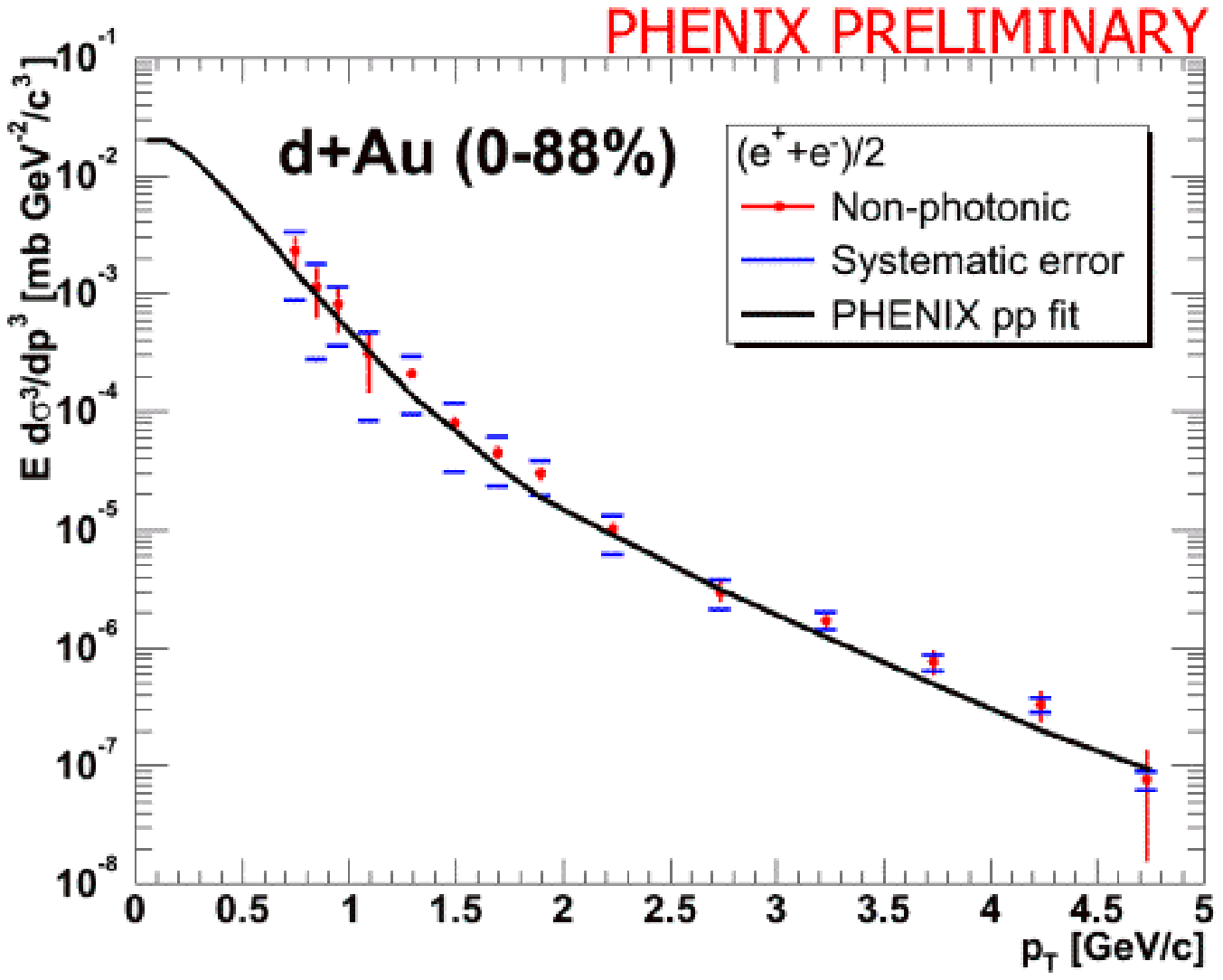}
\vspace*{-0.8cm}
\caption[]{
The non-photonic single electron spectra from the {\it d+Au} collisions
at $\sqrt{s_{NN}} = 200\ GeV$ (scaled 
with the number of colliding nucleon pairs).
The empirical parametrization of the {\it p+p} spectra 
is also displayed. 
}
\label{fig:dAu}
\end{figure}

Fig.~\ref{fig:dAu} shows both the non-photonic single electron production
in the {\it d+Au} collisions at $\sqrt{s_{NN}} = 200\ GeV$
scaled with the number of colliding nucleon pairs, 
described in section~\ref{sec:theory}, 
and the empirical parametrization of {\it p+p} production.
We observe that
the non-photonic single electron production 
in the {\it d+Au} collisions
scales with the number of colliding nucleon pairs.
Further study was made for four centrality classes.
Non-photonic single electron spectra 
obtained from each centrality class
are also consistent with the {\it p+p} spectra 
scaled by the nuclear thickness.
The observation is consistent with the picture of
non-photonic single electrons 
produced by point-like interactions.

\section{Conclusion}\label{sec:Conclusion}

We have discussed exciting theoretical and experimental aspects 
of heavy quark production 
with emphasis on quantities measurable by PHENIX.
Selected results from PHENIX were presented and 
we note key experimental observations.

a) The non-photonic single electron spectra from {\it p+p} collisions
at $\sqrt{s_{NN}} = 200\ GeV$ indicate
a possible excess over the NLO pQCD heavy quark prediction.

b) Within errors, 
the non-photonic single electron spectra 
obtained from {\it d+Au} collisions 
both 
in minimum bias 
or 
in various centrality classes
exhibits scaling with the number of colliding nucleon pairs
from the {\it p+p} spectra.
Hence, production process for non-photonic single electrons
is consistent with point-like interactions. 

\section*{Acknowledgment}

We thank the staff of the Collider-Accelerator and Physics Departments
at BNL for their vital contributions. We acknowledge supports 
from the Department of Energy and NSF (U.S.A.), MEXT and JSPS(Japan),
CNPq and FAPESP(Brazil), NSFC(china), CNRS-IN2P3 and CEA(France),
BMBF, DAAD, and AvH (Germany), OTKA(Hungary), DAE and DST(India),
ISF(Israel), KRF and CHEP(Korea), RMIST, RAS, and RMAE (Russia ),
VR and KAW (Sweden), U.S. CRDF for the FSU, US-Hungarian 
NSF-OTKA-MTA, and US-Israel BSF.
 

\vfill\eject
\end{document}